\documentclass{jpsj3}
\usepackage{txfonts,mathrsfs}
\usepackage{url}
\usepackage{xcolor}

\title{
Emergence of Elastic Softening Featuring Ultra-Slow Dynamics\\
Around Magnetic Critical Endpoint in UCoAl
}

\author{
Masahito Yoshizawa$^1$\thanks{yosizawa@iwate-u.ac.jp}, 
Yusei Shimizu$^2$,
Yoshiki Nakanishi$^1$, 
Yoshiya Homma$^2$, \\
Ai Nakamura$^2$, 
Fuminori Honda$^3$, 
and 
Dai Aoki$^2$
}

\inst{
$^1$Faculty of Science and Engineering, Iwate University, Morioka, Iwate 020-8551, Japan \\
$^2$Institute for Material Science, Tohoku University, Oarai, Ibaraki 311-1313, Japan\\
$^3$Central Inst. of Radioisotope Science \& Safety, Kyushu University, Fukuoka 819-0395, Japan\\
} 

\abst{
We conducted an investigation on the temperature and magnetic field dependence of the elastic properties of UCoAl. 
The longitudinal elastic stiffness, $C_{33}$, exhibits significant softening as the system approaches the critical endpoint (CEP). 
This softening is indicative of an ultrasonic dispersion phenomenon, where the anomaly in the elastic constants diminishes with increasing measurement frequency. 
Fine structures were observed near the CEP in higher frequencies. 
The magnetic field dependence of $C_{33}$ can be explained by assuming a specific field dependence of the relaxation time. 
Remarkably, we recorded a relaxation time of 3.5$\times$10$^{-8}$ s in the vicinity of the CEP, which is the longest observed value among solids.
These peculiar ultrasonic properties cannot be explained solely by Ising-like ferromagnetic fluctuations, 
 suggesting the involvement of the quadrupole (orbital) degree of freedom in the formation of the CEP. 
We discussed the origin of these observed phenomena in relation to the magnetic ground state of the UCoAl system.
}

\begin{document}
\maketitle

\section{Introduction}
As the temperature of many gases decreases, they undergo a phase transition and condense into liquids. 
At the boundary between the gas and liquid phases, the volume undergoes a discontinuous change. 
This phase boundary eventually reaches a critical endpoint (CEP) as pressure and temperature increase, where the volume changes continuously. 
At this CEP, a crossover state emerges, known as the critical liquid, in which the gas and liquid phases coexist harmoniously. 
This unique state, characterized by its high activity, finds extensive applications in industrial processes such as semiconductor cleaning and chemical reactions.

The CEP has captivated the attention of numerous researchers due to its intriguing nature. 
The Van der Waals equation, which was the first mathematical description of the CEP, states that the first derivative of pressure with respect to volume, which is closely related to the elastic stiffness constant, becomes zero at the CEP. 
This implies that matter loses its hardness at the CEP. In fact, a phenomenon has been observed in water where the sound velocity is significantly reduced at the CEP~\cite{Wagner2002}.
This softening of acoustic phonons and the corresponding decrease in sound velocity offer insights into unexplored phenomena and present a wide range of research opportunities.

The study of CEP phenomena in solids has been a subject of intensive research, particularly in connection with tri-critical point (TCP) phenomena and quantum phase transitions~\cite{Brando2016}.
In systems with valence fluctuations, where the order parameter is volume, ultrasonic studies have revealed effects similar to those observed in gaseous liquids. Large anomalies in longitudinal elastic constants have been observed in valence-fluctuating CeTh mixed crystals and YbIn$_{1-x}$Ag$_{x}$Cu$_{4}$~\cite{Wehr1981,Zherlitsyn1999}.
An example of a well-known case is the $\gamma$- and $\alpha$-Ce structural phase transitions under pressure, where a first-order transition associated with significant volume changes terminates at the CEP  ($P\sim 1.5\,{\rm GPa}$ and $T\sim 500\,{\rm K}$)\cite{Nikolaev2012}.
Furthermore, it is recognized that phase transitions in gaseous liquids and magnetic systems belong to the same universality class, with pressure and volume of the liquid being replaced by magnetization and magnetic field~\cite{Lee1952}, respectively. In magnetic systems, the magnetic susceptibility plays a crucial role instead of compressibility in liquids. 
As a result, magnetic CEPs have primarily been studied through magnetic measurements. However, it remains unclear how elastic properties manifest in magnetic CEPs.

\begin{figure}[ht]
\centering
\includegraphics[width=10cm]{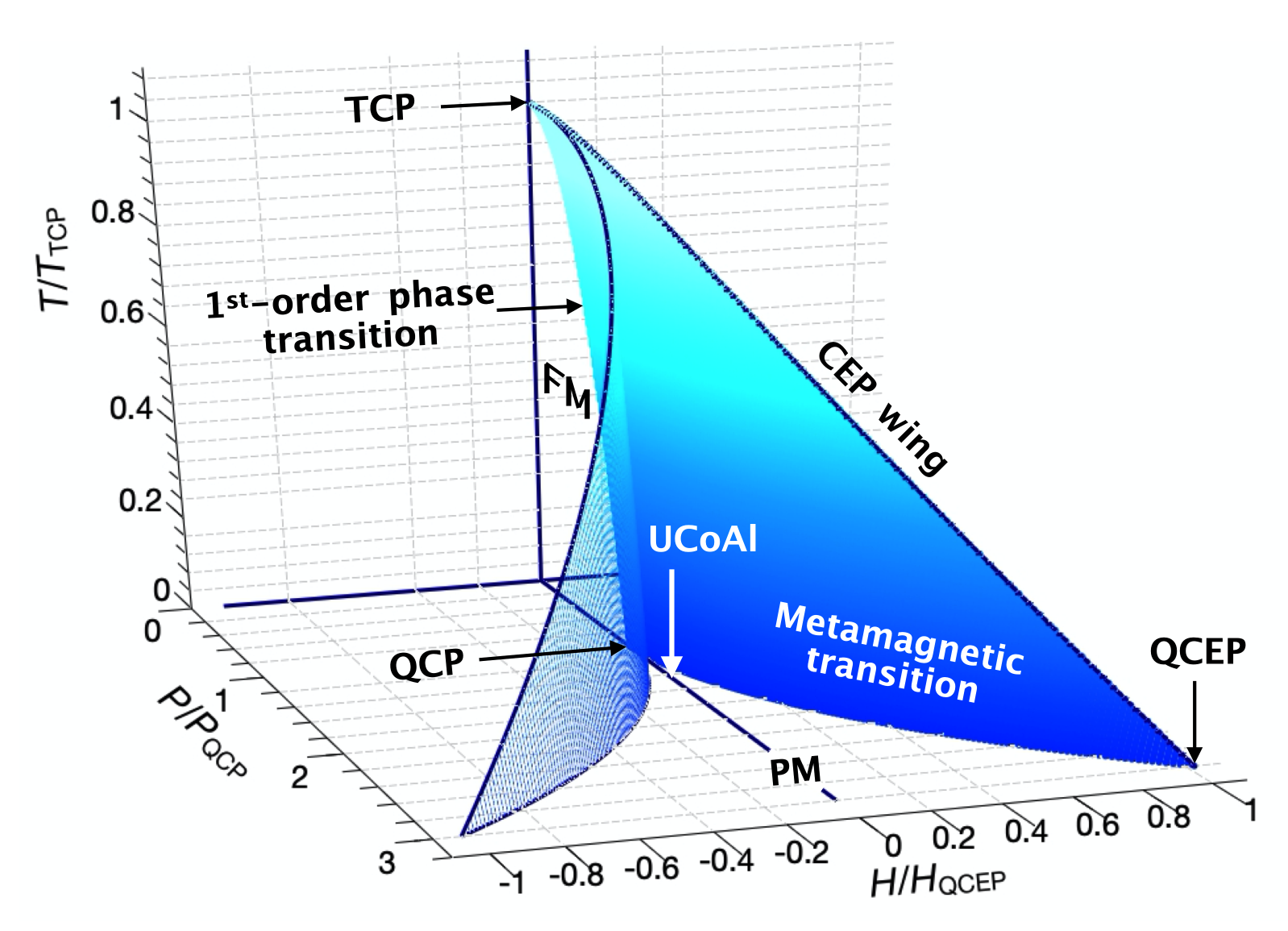}
\caption{
(Color online) Schematic magnetic systems illustrating the first-order phase transition as a function of temperature $T$, hydrostatic pressure $P$, and magnetic field $H$. The axes are normalized by the TCP temperature, the QCP pressure, and the field at QCEP, respectively. This diagram was generated by the application of Ren{\' e} Thom's method~\protect\cite{Thom1975} to Ginzburg--Landau free energy with reasonable parameters.
}
\label{fig1}
\end{figure}

Figure~\ref{fig1} is a schematic depicting the first-order phase transition in magnetic systems; the transition temperature $T$ is plotted as a function of pressure $P$ and magnetic field $H$. 
The first-order phase transition originates from a TCP, which marks the boundary between the second-order and first-order phase transitions. In this case, the temperature of the first-order transition decreases with increasing hydrostatic pressure $P$, and eventually disappears at a quantum critical point (QCP). 
The wings CEP originate from the TCP and terminate at a quantum critical endpoint (QCEP). 
The temperature, pressure, and field in Fig.~\ref{fig1} are normalized by the transition temperature at the TCP, the pressure at the QCP, and the field at the QCEP, respectively. Above the QCP, the system exhibits paramagnetic behavior at zero magnetic field and undergoes a metamagnetic transition upon the application of a magnetic field.

UCoAl is classified as an itinerant magnet and exhibits paramagnetic behavior in the absence of an applied magnetic field.
However, when a magnetic field is applied, a metamagnetic transition occurs at a critical field $H_{\text m}$. As the temperature increases, the abrupt metamagnetic transition gives way to a continuous transition at the CEP. Above this temperature, there exists a supercritical region and a crossover region. 
The CEP of UCoAl at ambient pressure has been reported to occur at approximately 11 K and 0.9 T. 
With increasing hydrostatic pressure, the CEP wing extends to the quantum critical endpoint (QCEP) at around 1.6 GPa~\cite{Aoki2011}. 
The estimated position of the QCP in UCoAl is -0.3 GPa~\cite{Mushnikov1999}. 
Consequently, UCoAl is considered to be in close proximity to the QCP and a ferromagnetic phase.

It is worth noting that UCoAl exhibits strong electronic correlations, with an electronic specific heat coefficient of 73 mJ/mol$\cdot$K$^{2}$~\cite{Matsuda1999}. 
Related materials, such as UGe$_{2}$, have been found to exhibit superconductivity in the vicinity of the ferromagnetic QCP and QCEP~\cite{Taufour2010,Kotegawa2011}. 
The interplay between the superconducting mechanism, ferromagnetic correlations, and the critical points in these materials has garnered significant attention. UCoAl manifests ferromagnetism under uniaxial pressure or chemical substitution~\cite{Shimizu2015}. 
Due to its diverse range of physical properties dependent on external parameters, UCoAl has been extensively investigated by researchers as a representative material for elucidating the physical properties near the TCP and CEP~\cite{Combier2016}. 
Given its analogous characteristics, UCoAl is considered to be a suitable material for such studies.

The $5f$ electrons of uranium (U) exhibit dual characteristics, combining aspects of both localized and itinerant behavior. 
In the context of ultrasonic investigations, the localized character manifests in the transverse elastic constants through symmetry changes, while the itinerant character is evident in the longitudinal elastic constants with local volume changes. 
The primary focus of this study is the observation that the crystal symmetry of UCoAl remains largely unchanged at CEPs, except for the alterations associated with the magnetic transition. By employing precise measurements of the longitudinal elastic constants $C_{33}$ using the elastic strain $\epsilon_{zz}$ as a probing technique, we aim to uncover the distinctive properties of the critical liquid state. 
Through our ultrasonic measurements, we will present intriguing elastic properties of UCoAl, shedding light on its unique characteristics.

\section{Experiment}
Single crystals of UCoAl were grown using the Czochralski method in a tetra-arc furnace. 
Two single crystals were selected for measurements. 
The dimensions of the first crystal were 1.15($a$) $\times$ 1.0($b$) $\times$ 1.936($c$) mm$^{3}$, while the second crystal had dimensions of 1.09($a$) $\times$ 0.5($b$) $\times$ 1.338($c$) mm$^{3}$.Here, $a$, $b$, and $c$ represent the $[10\bar{1}0]$, $[11\bar{2}0]$, and $[0001]$ directions, respectively, in the hexagonal crystal structure.
To calculate the density $\rho$ of UCoAl, the lattice constant was used, resulting in a value of 10.5 g/cm$^3$.

Ultrasonic measurements were conducted using a measurement system that combines the DynaCool instrument from Quantum Design and the Ultrasonic Option from Quantum Design Japan (available at \url{https://www.qd-japan.com/products/ppms-ultrasonic-elastic-constant-measurement-option/}).
This system allows for simultaneous measurement of ultrasonic sound velocity and ultrasonic attenuation in a temperature range from 300 to 2 K and a magnetic field range from -9 to 9 T.
In this study, we employed the phase-quadrature method with a fixed measurement frequency, known as the ORPHEUS method~\cite{Fujita2010}.
Measurements were carried out using five different frequencies: 12.10, 34.25, 60.92, 85.61, and 106.5 MHz, which are abbreviated as 12, 34, 61, 86, and 106 MHz, respectively.

LiNbO$_3$ piezoelectric transducers were utilized for sound wave emission and detection in the ultrasonic measurements. 
The transducers had $Z$-cut and $X$-cut orientations, with fundamental frequencies of 12 and 36 MHz for longitudinal waves, and 18 MHz for transverse waves.
For the 34 and 106 MHz measurements, 36 MHz transducers were employed, while 12 MHz transducers were used for the 12, 61, and 85 MHz experiments.
The value of the longitudinal elastic constant $C_{33}$ was determined from the longitudinal sound velocity $v_{33}$ propagating along the $c$-axis, using the equation $C_{33} = \rho v_{33}^2$, where $\rho$ represents the density of the material. 
Similarly, the other elastic constants, $C_{11}$ and $C_{44}$, were obtained from the corresponding sound velocities. 
Specifically, $C_{11}$ was obtained from the velocity of the longitudinal sound wave propagating along the $a$-axis, while $C_{44}$ was derived from the transverse sound wave traveling along the $c$-axis with the displacement along the $a$-axis or $b$-axis, respectively.

The echo signals for UCoAl can be found in the Supplemental material~\cite{supplUCoAl2023}. 
To obtain the absolute values of sound velocity, a comparison between the results from two samples was conducted. The echo distance and frequency change for a 360-degree phase change were used to determine these values.
The accuracy of the measured elastic constants is approximately $\pm$ 5 \%.
Since UCoAl possesses a hexagonal crystal structure, the value of $C_{12}$ was obtained using the equation $C_{66} = (C_{11} - C_{12})/2$, along with the experimental value of $C_{11}$.
In this study, measurements of the temperature and magnetic field dependencies of $C_{33}$ and $C_{44}$ were carried out on a sample with a length of 1.936 mm along the $c$-axis. 
While ultrasonic attenuation measurements were also performed, this paper focuses primarily on the elastic constants for the reasons explained later.

\begin{table}[ht]
\centering
\begin{tabular}{cccc}
\hline
 Elastic constant & Sound Velocity (km/s)  & Value (GPa) \\
\hline
\hline
$C_{11}$ & 3.63 & 138    \\
$C_{33}$ & 4.08 & 175    \\
$C_{44}$  & 2.46 &  63.5 \\
$C_{66}$ & 2.02 &  42.8  \\
$C_{12}$ &  & 52.7   \\
\hline
\end{tabular}
\caption{\label{table1}Elastic stiffness constants of UCoAl at 200 K.}
\end{table}

\section{Results and Analyses}
\subsection{Temperature and Magnetic Field Dependences of $C_{33}$}
\begin{figure}
\centering
\includegraphics[width=12cm]{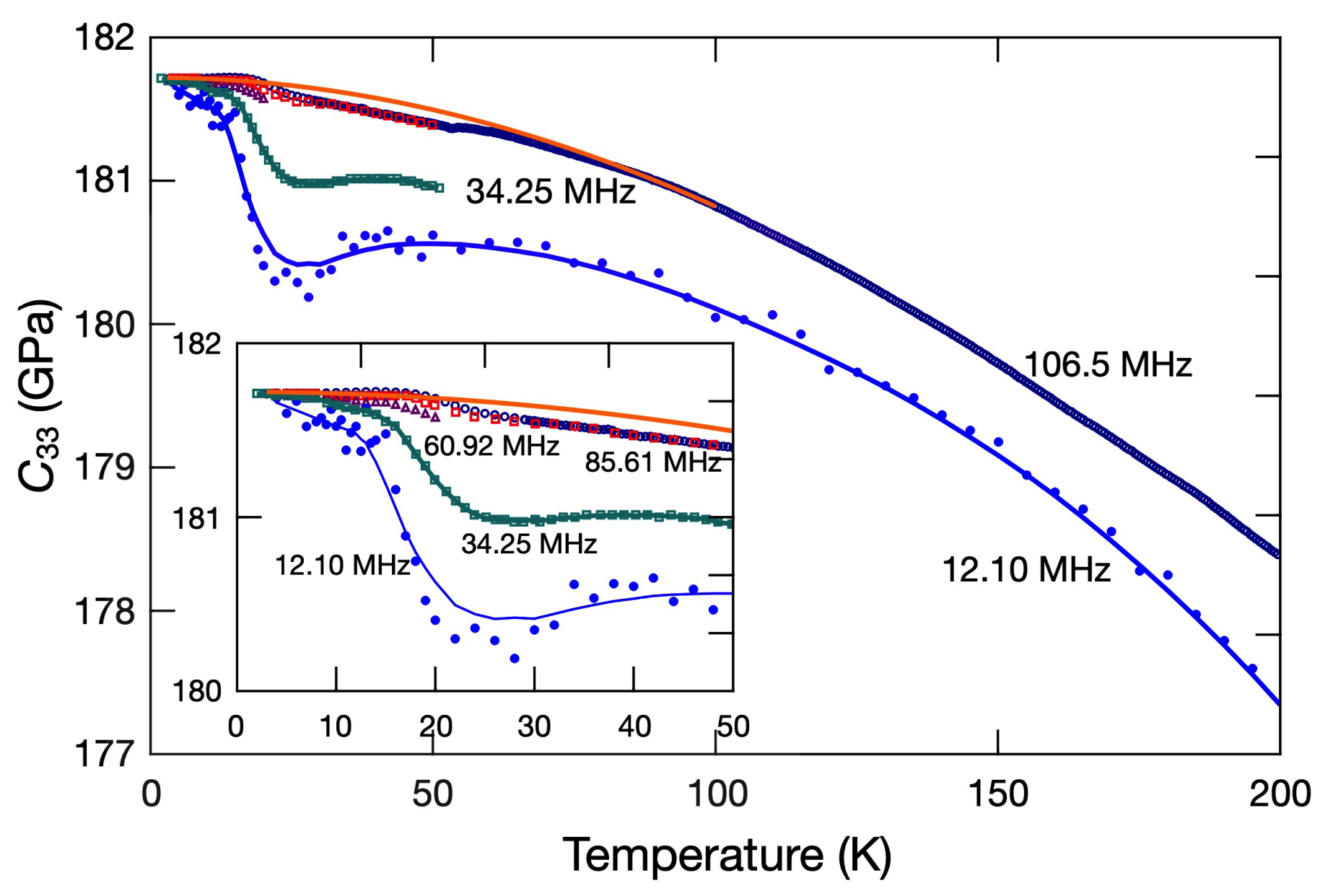}
\caption{
(Color online) Temperature dependence of $C_{33}$ measured at five frequencies. 
The thick orange line in Figure~\ref{fig2} represents the background contribution arising from lattice anharmonicity. 
This background is described by the equation $C_{\rm 33back} = 181.7 - 9.00\times10^{-5}T^{2}$. The inset shows the detailed behavior of $C_{33}$ below 50 K.
}
\label{fig2}
\end{figure}

Figure~\ref{fig2} illustrates the temperature dependence of $C_{33}$ below 200 K, measured at five different frequencies: 12, 34, 61, 86, and 106 MHz. 
The values of elastic constants at 200 K for UCoAl are provided in Table~\ref{table1}. 
To facilitate comparison, the data for each frequency are normalized by the values at low temperature in zero magnetic field.
Remarkably, distinct anomalies are observed at low frequencies of 12 and 34 MHz at 20 K. 
As the frequency increases, these anomalies diminish, and the temperature range over which they appear significantly narrows at 106 MHz. 
This indicates that $C_{33}$ exhibits ultrasonic dispersion, where the elastic constants are frequency-dependent.
The orange curve represents the lattice contribution, which is obtained from the data at 106 MHz.This background curve is utilized to extract the anomalous part of the temperature variation of $C_{33}$, referred to as $C_{33}( \infty )$. It is assumed that $C_{33}(\infty)$ remains independent of the magnetic field.

\begin{figure}
\centering
\includegraphics[width=12cm]{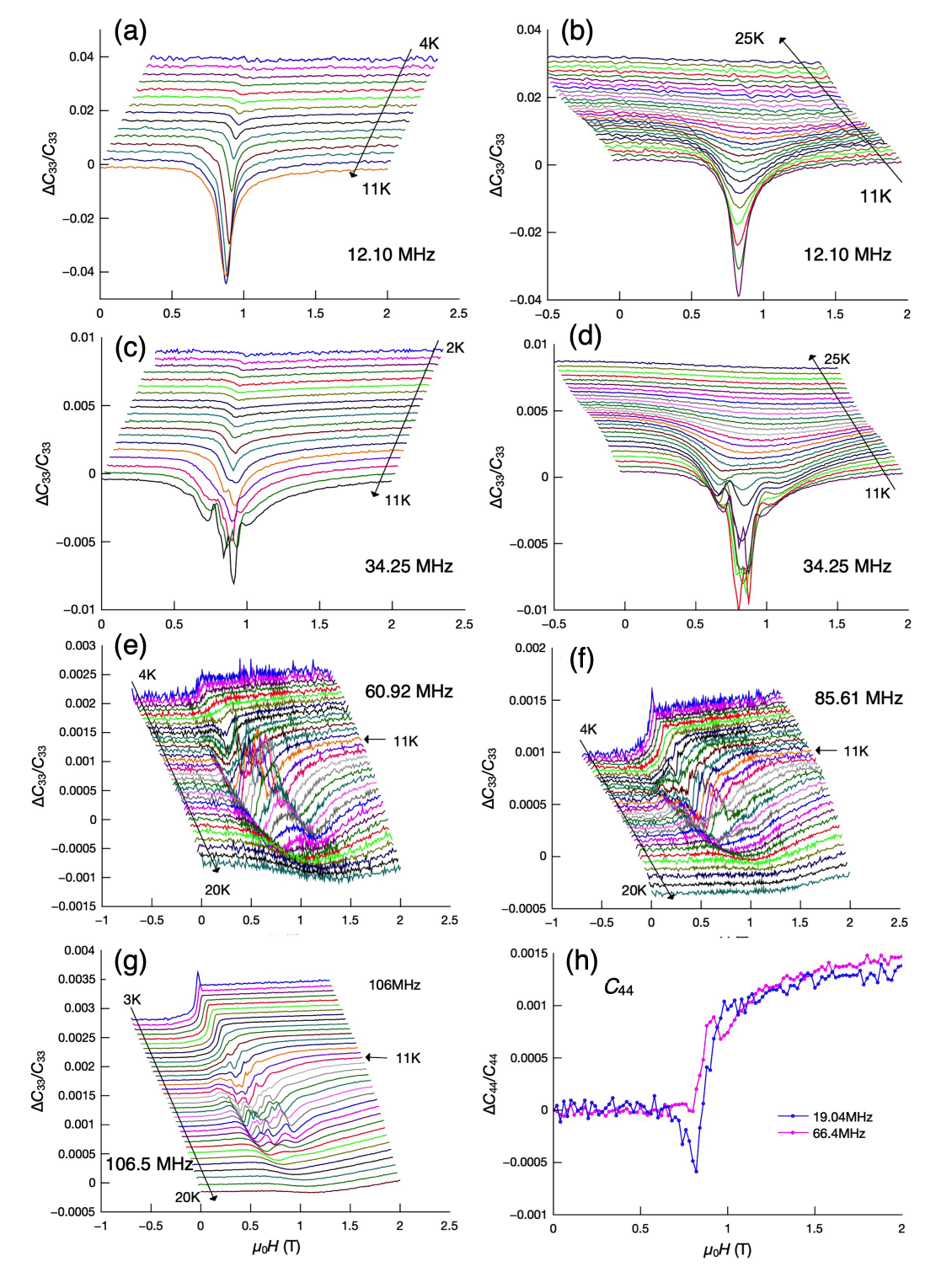}
\caption{
(Color online) Water fall description of magnetic field dependence of $C_{33}$ under various measurement frequencies ranging from 12MHz to 106 MHz. 
The magnetic field dependence of $C_{44}$ at 19 and 66 MHz are also plotted.
}
\label{fig3}
\end{figure}

Figure~\ref{fig3} presents the anomalous parts of $C_{33}$ in a waterfall plot, showing the magnetic field dependence. 
The measurements were performed by incrementally increasing the magnetic field from 2 to 20K, with a step of 1K, and holding the field at every 100 Oe. 
The data were interpolated using the Akima spline method at 0.5 K and every 50 Oe.
At 12 and 34 MHz, $C_{33}$ exhibits elastic softening toward the CEP. 
The anomaly at the CEP is more pronounced in the 12 MHz measurement, with an approximately 5\% decrease in $C_{33}$. 
The 34 MHz data also display a clear elastic softening near the CEP, although the magnitude of the anomaly is smaller compared to the 12 MHz data. 
In addition, at 34 MHz, a fine and oscillating field-dependent structure is observed near the CEP.

At 61 MHz, the softening at the CEP is suppressed, and a maximum is observed instead. 
The anomaly at the CEP almost disappears at 86 and 106 MHz. 
However, at these higher frequencies, a finely oscillating behavior of the elastic constants is still observed near the CEP, similar to the behavior observed at 34 MHz. 
Moreover, at frequencies above 61 MHz, a slight increase in the elastic constants is observed in the high-field region beyond the metamagnetic transition. 
This indicates a slight but significant stiffening of the material in the high-field ordered phase. 
In contrast, no such change is observed at 12 and 34 MHz due to the significant elastic softening near the metamagnetic transition.

For comparison, the magnetic field dependence of $C_{44}$ is shown at 19 and 66 MHz. At 19 MHz, a small elastic anomaly is observed in $C_{44}$ at the CEP, although it is not as prominent as in $C_{33}$. 
The anomaly at 19 MHz disappears at 66 MHz, suggesting the presence of ultrasonic dispersion in $C_{44}$ as well.
The elastic softening and ultrasonic dispersion phenomena near the CEP are considered to be more pronounced in the longitudinal elastic constants ($C_{33}$) rather than the transverse elastic constants ($C_{44}$).

\begin{figure}
\centering
\includegraphics[width=14cm]{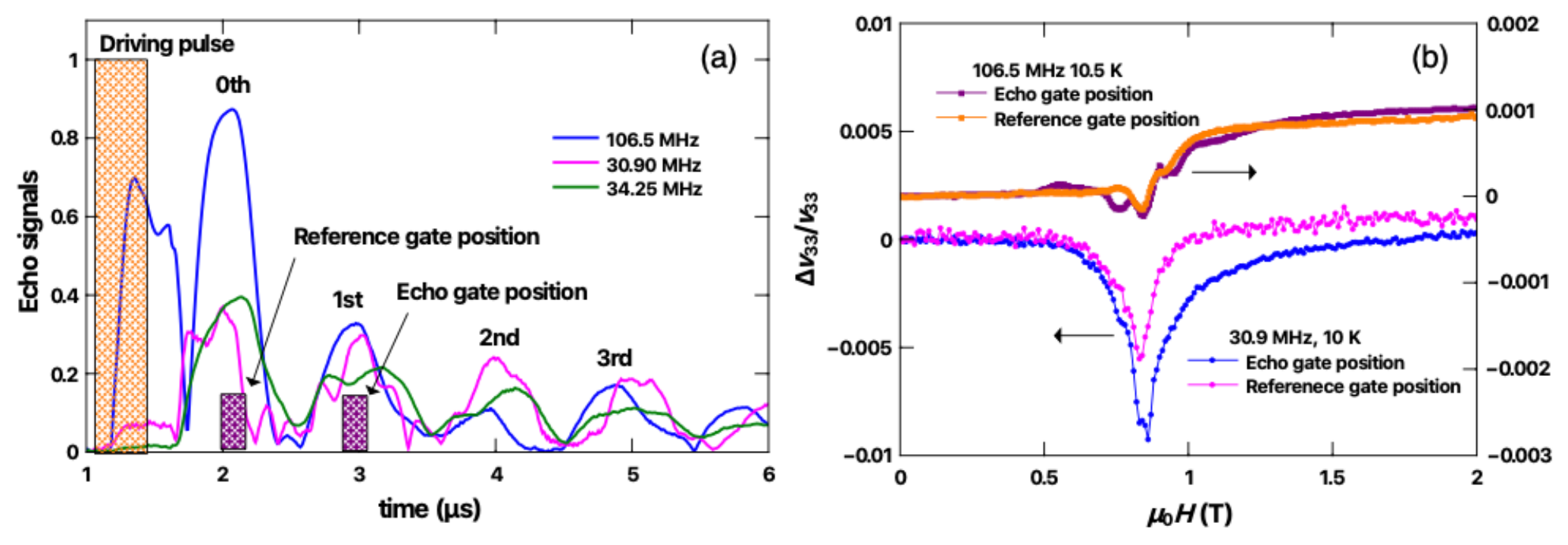}
\caption{
(Color online) (a) Echo trains of the $C_{33}$ mode recorded for frequencies of 30.90, 34.25, and 106.0 MHz as a function of time. 
The positions of the driving pulse, Echo gate, and Reference gate are indicated.
(b) Magnetic field dependence of the velocity change for the $C_{33}$ mode, measured at the Echo gate position for 30.90 MHz and at the Reference gate position for 106.0 MHz.
}
\label{fig4}
\end{figure}

\subsection{Experimental Evidence of Ultrasonic Dispersion}
Temperature and field dependences of $C_{33}$ show remarkable frequency dependence, which we believe provides clear evidence for the presence of ultrasonic dispersion in UCoAl. 
In a previous study by Kumano $et\ al.$ measured, they observed frequency-dependent anomalies in the temperature dependence of the $C_{44}$ elastic constant in the ferromagnet UCo$_{1-x}$Os$_{x}$Al~\cite{Kumano2020}. 
The temperature dependence of $C_{44}$ exhibited similarities to that of $C_{33}$ observed in our study, suggesting a common origin for the anomalies in $C_{33}$ and $C_{44}$.
Hence, it is not surprising that $C_{33}$ also displays frequency dependence.
In our experimental setup, we employed the ORPHEUS method. 
This method allows us to set two gate positions, namely the ``Echo gate position'' and the ``Reference gate position'' for detecting echo signals. 
The signals at the Echo gate position are used to calculate the velocity change, while the signals at the Reference gate position are used to determine ultrasonic attenuation, which is proportional to the logarithm of the echo amplitude ratio between the two gate positions. 
Figure~\ref{fig4}(a) illustrates the positions of the Echo gate and Reference gate relative to the driving pulse and the echoes. 
The Echo gate and Reference gate function independently for velocity measurements, unlike in a phase comparison method. 
In addition, the velocity change can be simultaneously obtained using the data at the Reference gate position through calculation. 
The coefficient for the 0$^{\text{th}}$ echo is three times larger than that for the 1$^{\text{st}}$ echo, as the effective sample length at the Reference gate position is three times longer than at the Echo gate position.
Figure~\ref{fig4}(a) presents the echo trains of $C_{33}$ at frequencies of 30.90, 34.23, and 106.5 MHz. The echo signals at each frequency appear at the same position as a function of time.
These results confirm that we measured the $C_{33}$ mode across the entire frequency range. 

Figure~\ref{fig4}(b) depicts the field dependence of the velocity change at the Echo gate position and Reference gate position for 30.9 and 106.5 MHz. Since we did not obtain data at the Reference gate position for 34.25 MHz, we present the data for 30.90 MHz instead. 
The data at 106.5 MHz only exhibit a step-wise anomaly above 0.9 T, while the data at 30.90 MHz clearly show softening toward $H_{\text m}$.
Moreover, the magnitude of the anomaly at 106.5 MHz is significantly smaller than that at 30.90 MHz. 
Although the data at the Echo and Reference positions for 106.5 MHz are nearly identical, except for a fine structure near $H_{\text m}$.
Slight differences exist between the data at the Echo gate position and the Reference gate position. 
We acknowledge that the anomaly around the CEP is frequency and gate position-dependent, as demonstrated in Fig.~\ref{fig3}. 
The field dependence of $C_{33}$ exhibits intricate fine structures around the CEP that cannot be solely explained by ultrasonic dispersion. 
The exact reasons behind these differences and fine structures are not yet clear and may be due to interference between neighboring echoes or other intrinsic factors. 
Nevertheless, our conclusion remains that the contrasting behaviors between 30.90 and 106.5 MHz provide compelling evidence for the presence of ultrasonic dispersion in UCoAl.

\subsection{Power-law Analysis of $C_{33}$}
\label{Powerlaw}
\begin{figure}
\centering
\includegraphics[width=15cm]{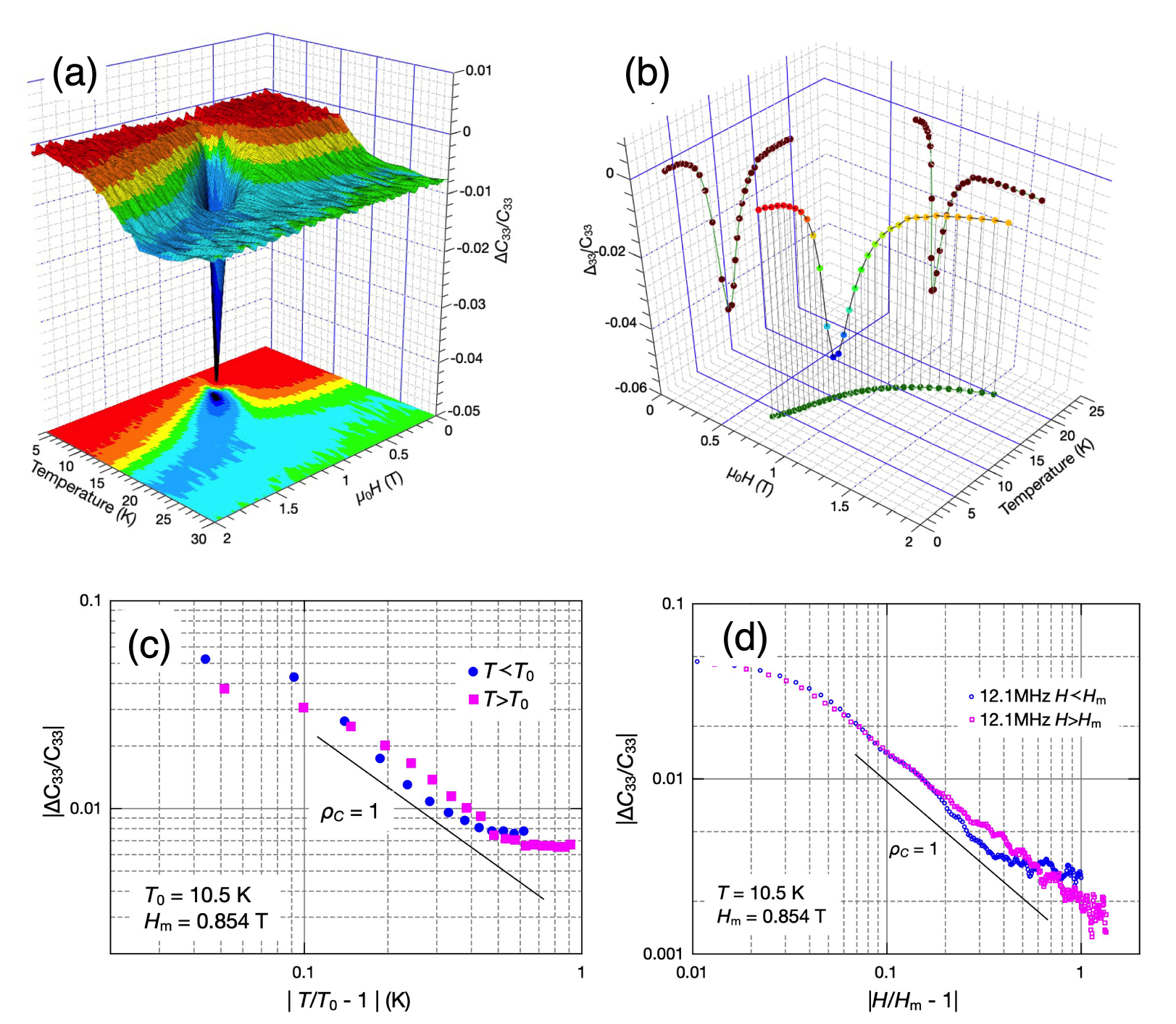}
\caption{
(Color online) (a) Relative variation of the longitudinal elastic constant $C_{33}$ as a function of temperature and magnetic field.
(b) Trace of $C_{33}$ anomalies as a function of temperature and magnetic field.
(c) Temperature dependence of the $C_{33}$ anomaly plotted on a logarithmic scale, obtained by projecting graph (b).
(d) Magnetic field dependence of the $C_{33}$ anomaly plotted on a logarithmic scale, obtained from magnetic field scan data.
}
\label{fig5}
\end{figure}

Figure~\ref{fig5}(a) displays the magnetic field-temperature dependence of $C_{33}$ at 12 MHz obtained in this study, revealing a pronounced softening in the vicinity of the CEP. 
Upon closer inspection of the 12 MHz $C_{33}$ data, it exhibits divergent behavior near the phase transition temperature $T_{\text{c}}$ following a power law of the form $\Delta C = (T - T_{\text c})^{-\rho_{C}}$. 
In systems where the order parameter and the elastic strain are coupled through magnetostrictive interactions, the critical exponent $\rho_{C}$ can be expressed as $\rho_{C} = \alpha + 2\phi - 2$, where $\alpha$ is the critical exponent of specific heat~\cite{yoshizawa1982}. 
For elastic modes that do not change the symmetry of the system, $\phi$ is equal to 1, while for transverse elastic modes that induce a symmetry change, $\phi$ takes a different value. 
In the case of $C_{33}$, we expect $\rho_{C} = \alpha$ since it corresponds to an elastic mode that does not alter the system's symmetry.

To determine the critical exponents of $C_{33}$ for the magnetic field dependence and temperature dependence, we followed the following procedure. 
Figure~\ref{fig5}(b) shows a three-dimensional graph of the temperature-field profile, illustrating the maximum anomalous amount of elastic softening for $C_{33}$.
By projecting the data from Fig.~\ref{fig5}(b) onto the temperature axis, we obtained the temperature dependence of the anomalous part of $C_{33}$, which is displayed logarithmically in Fig.~\ref{fig5}(c). 
For the magnetic field dependence, we performed measurements at 10.5 K, as shown in Fig.~\ref{fig5}(d). 
The power-law analysis reveals that both the magnetic field dependence and the temperature dependence diverge toward the CEP with a critical exponent of 1. 
However, previous studies have supported the three-dimensional (3D) Ising model results, which predict a different critical exponent for UCoAl.
In the 3D Ising model, the value of $\alpha$ is 0.11~\cite{Guillou1980}. 
The observed value of $\rho_{C}$ does not align with that of magnetostrictive coupling, indicating that the elastic anomaly is caused by Jahn-Teller-type interactions, where the strain linearly couples with an unknown order parameter. 
Plausible candidates for the order parameter include orbital, electric quadrupole, and charge degrees of freedom in U atoms and a nematic order parameter formed by magnetic dipoles. In this study, we will refer to the unknown order parameter as ``quadrupole'' since the $Q_{zz}$ quadrupole possesses the same symmetry as the $\epsilon_{zz}$ strain. 
The Jahn-Teller energy $E_{\text{JT}}$, which is defined as the proportionality constant in the power law with an exponent of 1, is found to be 0.042 K for $T > T_{0}$ and 0.033 K for $T < T_{0}$.

\subsection{Ultrasonic Dispersion}
The frequency dependence of the elastic constant is characterized by two values: $C(0)$ and $C( \infty )$. Here, $C(0)$ represents the static or low-frequency value of the elastic constant, while $C(\infty)$ represents the background value of the elastic constant. 
These values provide important information about the frequency dependence and behavior of the elastic constant in the system.

\begin{eqnarray}
C \left ( \omega \right )  =C \left ( \infty \right )  - \frac {C \left ( \infty \right )  - C \left ( 0\right )  } { 1 + \left ( \omega \tau \right )^{2}  } \\
\alpha \left ( \omega \right )  = \frac { C \left ( \infty \right )  - C \left ( 0 \right )  } { 2 \rho v^{3} } \frac { \omega^{2} \tau }{ 1 + \left ( \omega \tau \right )^{2}  } 
\end{eqnarray}

In this measurement, the evaluation of relaxation time using simultaneous measurements of sound velocity and attenuation coefficient was not successful due to the large ultrasonic attenuation observed in the system. 
The equation commonly used for ultrasonic absorption, which relates relaxation time to attenuation, may not be applicable when the relaxation time is significantly slow.
In certain studies of ultrasonic dispersion of liquids with relaxation times comparable to those observed in this study, the analysis was performed by Cole-Cole plots, which are used in dielectric dispersion, instead of the above equation~\cite{Eden1973}.
Therefore, in this study, relaxation times were obtained indirectly from the elastic constants at different frequencies. 
The magnetic field dependence of the relaxation time was determined using the 12 and 34 MHz data at each temperature, employing a process described by equation (1) (hereafter referred to as ``Process''). 
These data exhibited a systematic behavior, allowing the construction of a model for the magnetic field dependence of the relaxation time at each temperature. 
The details of this model construction can be found in the Supplemental material~\cite{supplUCoAl2023}.
By utilizing this model, the elastic constants at each frequency were calculated based on the 12 MHz data. 
To assess the validity of the model, a comparison was made between the calculated values and the experimental data. 
This approach provided a way to estimate the relaxation time indirectly and evaluate its magnetic field dependence in the system.

\begin{figure}
\centering
\includegraphics[width=15cm]{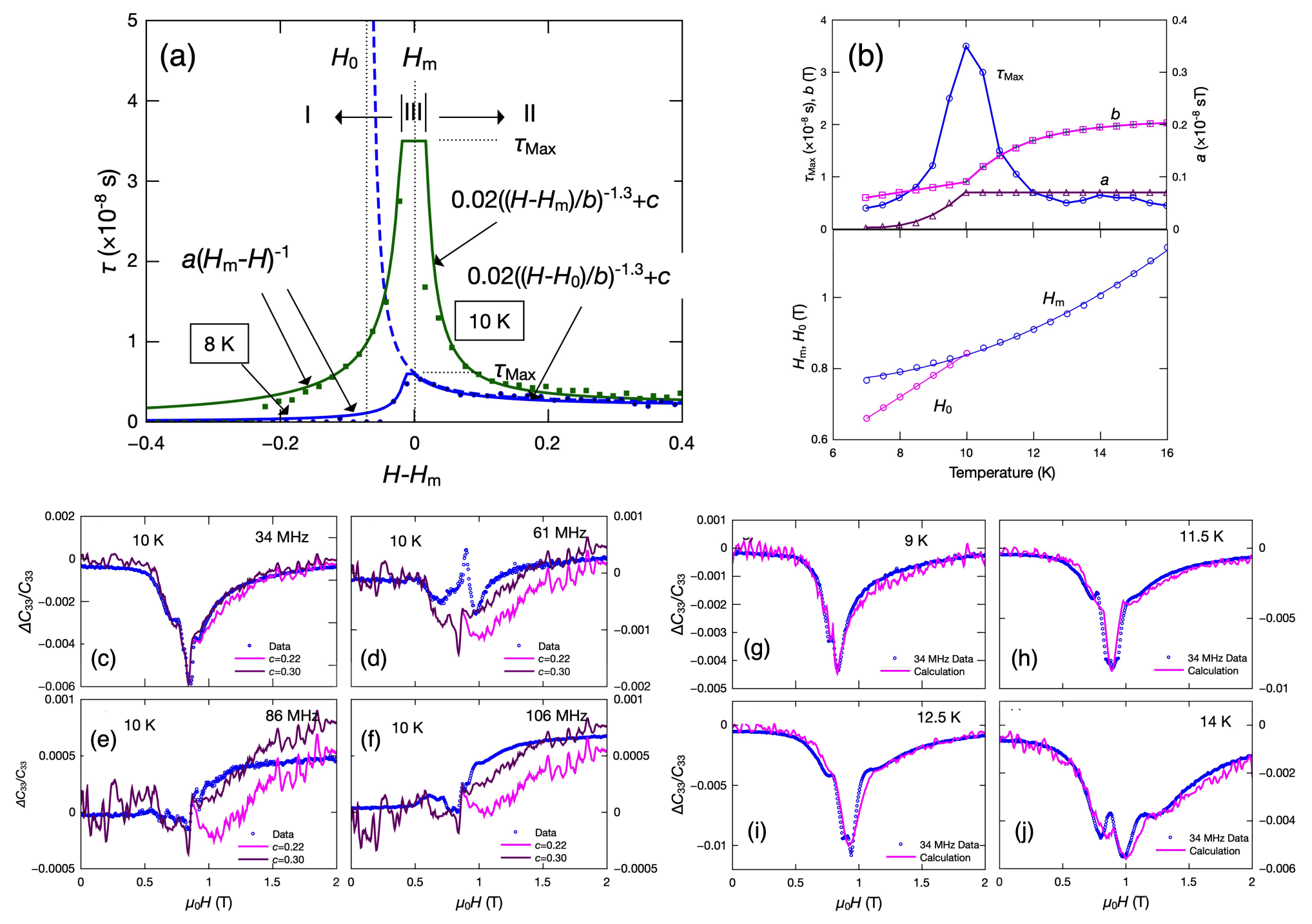}
\caption{
(Color online) (a) Model of relaxation time as a function of magnetic field consisting of two parts: scaling behavior over a wide range of field strengths (labeled as I and II) and an insensitive behavior to magnetic field near the metamagnetic transition point ($H_{\text m}$).
(b) Temperature dependence of parameters $a$, $b$ and $\tau _{\text bMax}$ (upper), and $H_{\text m}$ and $H_{0}$ (lower).
Magnetic field dependences of the experimental data are plotted alongside the calculations performed using the 12 MHz data and the relaxation time. 
Results for 34, 61, 86 MHz, and 106 MHz data are shown in (c)--(f). 
Results for 34 MHz data at selected temperatures are shown in (g)--(j).
}
\label{fig6}
\end{figure}

Figure~\ref{fig6}(a) illustrates the model derived from the magnetic field dependence of the relaxation time. 
The plot includes relaxation times obtained from 12 and 34 MHz data at 10 K. It is evident that the relaxation time increases as the magnetic field moves from low values toward $H_{\text m}$ and also from high magnetic field values, following a power law. 
For the low field side, an exponent of 1 was used, while for the high-field side, an exponent of 1.3 was employed. 
However, to accurately represent the data near $H_{\text m}$, it was necessary to strongly suppress this divergent behavior. 
The regions that conform to the power-law behavior are collectively referred to as Regions I and II, while regions where $\tau$ remains constant or exhibits weak field dependence are collectively referred to as Region III. This characteristic behavior was consistently observed in the temperature region above the CEP. 
In the proposed model, the relaxation time in Region III is assumed to be constant.
Next, we discuss the parameters of this model.
In the temperature region above 10 K, the relaxation time obtained from the elastic constant data in Region II can be described by the equation $0.02 \left( \frac{H - H_{\text{m}}}{b} \right)^{-1.3} + c$. Here, $H_{\text{m}}$ is determined from the locus of the minima in the elastic constants at 12 MHz. 
The parameter $c$ is set to 0.22 $\times$ 10$^{-8}$ s, which was obtained from the process. In the metamagnetic region below 10 K, the same equation is used due to the observed scaling law. However, in this case, the power-law behavior diverges toward the low magnetic field $H_{0}$ instead of $H_{\text{m}}$. 
The power-law behavior in Region I is described by the equation $a \left(H_{\text{m}} - H\right)^{-1}$, where $a$ is a parameter.
Above 10 K, the parameter $a$ remains constant and is fixed at 0.07. In the temperature range from 8-10 K, the relaxation time seems to follow a power law, and an appropriate value of $a$ 
is chosen to fit the data. At lower temperatures, the relaxation time diverges toward $H_{0}$ at low magnetic fields. 
Interestingly, even at lower temperatures, there was no significant difference in the reproducibility of the data, whether the relaxation time follows a power law or changes toward a staircase behavior at $H_{\text m}$. 
Therefore, to maintain analytical continuity, the same power-law representation was adopted for temperatures higher than the CEP.
The maximum relaxation time, $\tau_{\text Max}$, was determined by fitting the 34 MHz data using this model along with the 12 MHz data. 
The temperature dependence of the parameters mentioned above is summarized in Fig.~\ref{fig6}(b).

Figures~\ref{fig6}(c)--(f) present the results obtained at 10 K for each frequency, calculated using the 12 MHz data and the proposed model, alongside the measured values. 
The 106 MHz data were used as $C_{\infty }$.
Notably, the 34 MHz data exhibit slight up-down behavior near the CEP. In this model, these anomalies are attributed to the boundaries between Regions I to III and II to I. 
Figure \ref{fig6}(g) to (j) specifically focus on the calculations and data at 34 MHz for selected temperatures, while the results for other temperatures are provided in the Supplemental material~\cite{supplUCoAl2023}. In all cases, the fine field dependence of the data near $H_{\text{m}}$ is accurately reproduced. In Figs.~\ref{fig6}(c)--(f), we adopted $c$ to be 0.22 $\times$ 10$^{-8}$ s  and 0.30 $\times$ 10$^{-8}$ s.
The former value was obtained from the process. 
It is evident that a slight adjustment of the value of $c$ improves the fitting. 
Although we are uncertain about the factors contributing to $c$, it is crucial to emphasize that fitting without $c$ is impossible. The field-dependent behavior of $C_{33}$ at 34 MHz is well captured by assuming the model discussed above, which considers the field dependence of the relaxation time. 
However, reproducing all the details at higher frequencies was not feasible. In particular, the data at 61 MHz exhibit an elastic constant maximum at $H_{\text m}$. 
We speculate that the fine structure of this behavior, as a function of the field, cannot be solely explained by ultrasonic dispersion, and some unknown factors contribute to such observations. Nonetheless, the overall features can be accounted for by the ultrasonic dispersion effect, indicating the fundamental validity of the proposed model.

\begin{figure}
\centering
\includegraphics[width=15cm]{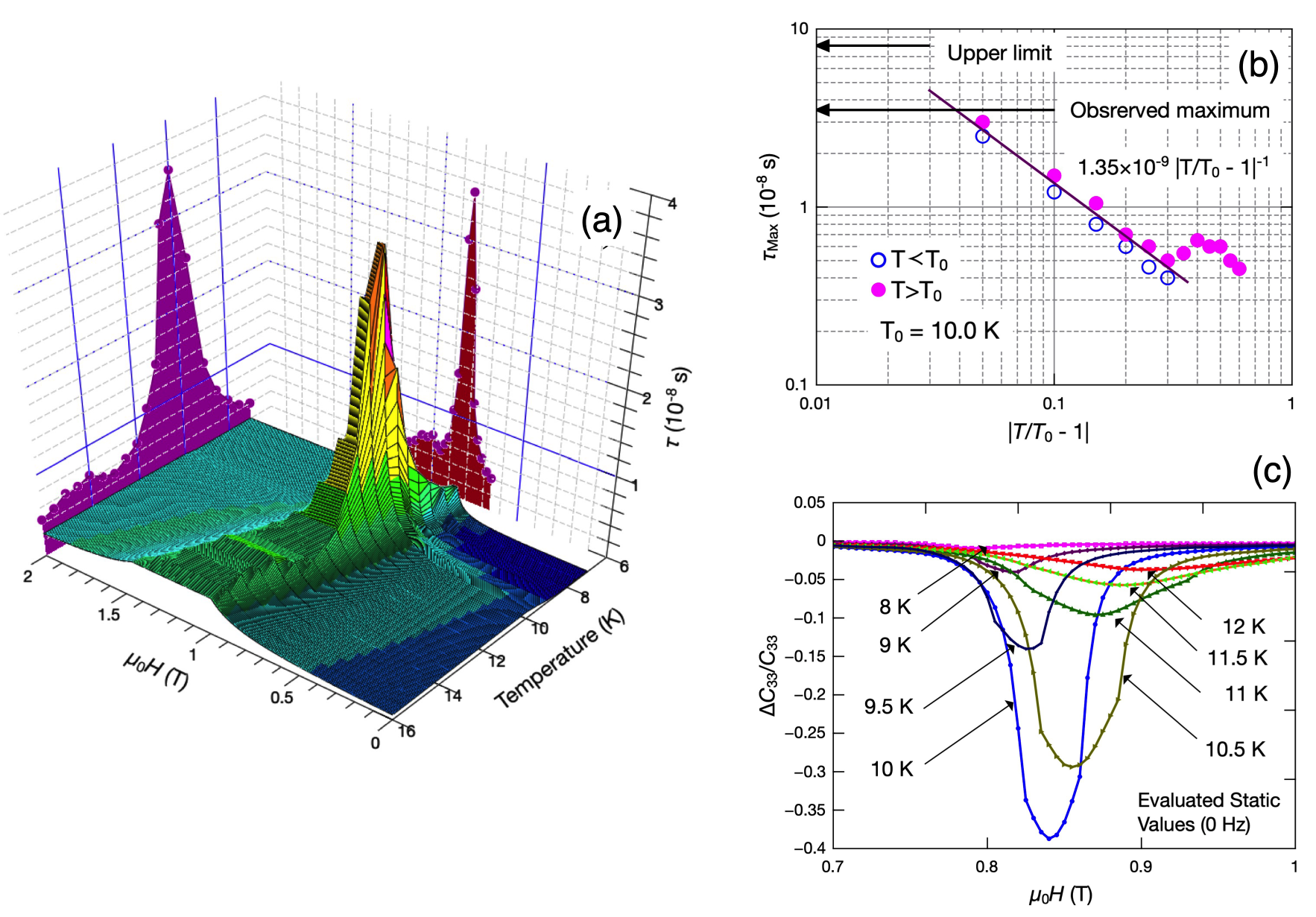}
\caption{
(Color online) (a) 3D graph of relaxation time as a function of temperature and magnetic field.
(b) Relaxation time values for Region III as a function of temperature on a logarithmic scale.
Data above and below CEP follow a power law with the critical exponent set to 1.
(c) Magnetic field dependence of the static limit of $C_{33}$ calculated using relaxation time values and 12 MHz data.
}
\label{fig7}
\end{figure}

Figure~\ref{fig7}(a) illustrates the temperature and magnetic field dependence of the relaxation time. 
Furthermore, Figure~\ref{fig7}(b) displays the trajectory of the maximum relaxation time, $\tau_{\text Max}$, as a function of temperature in logarithmic scale. 
Notably, $\tau_{\text Max}$  follows a power law with a critical exponent of 1, as evident from the figure.
Regarding the peculiar behavior of the field dependence of $\tau$, which is significantly suppressed in Region III, one plausible explanation is that $\tau$ consists of two components: a magnetic field-dependent component and a less sensitive component. 
The former is attributed to the magnetic contributions such as spin, dipole, and orbital effects, while the latter arises from the nonmagnetic components such as charge and electric quadrupole effects. 
These intriguing behaviors merit further investigation in future studies to gain a comprehensive understanding.

\section{Discussions}
\subsection{Ultra-Slow Dynamics and Possible Giant Elastic Softening}
In this section, we will demonstrate that a significant elastic anomaly is expected in $C_{33}$. 
It is important to note the ``upper limit'' depicted in Fig.~\ref{fig7}(b). 
This upper limit was determined through the following procedure: employing Equation 1, a maximum anomaly of approximately 5\% at 12 MHz was assumed, which led to the evaluation of the relaxation time value at which the static elastic constant $C(0)$ becomes zero as 8 $\times$ 10$^{-8}$ s. 
The upper limit signifies that if the relaxation time exceeds this value, $C(0)$ will become negative, indicating a breakdown in the structural stability of the system.
Figure~\ref{fig7}(c) displays the magnetic field dependence of the extrapolated anomalous part of $C_{33}$ to zero at the measuring frequency, as obtained from the calculations. 
The observed $\tau_{\text Max}$ was determined to be 3.5 $\times$ 10$^{-8}$ s, suggesting a predicted 40 \% elastic softening for $C(0)$ in the vicinity of the CEP. The CEP represents a singularity where the structural stability of the system collapses.

The universality class of the TCP wings in UCoAl is believed to be described by a 3D Ising model due to the presence of a magnetic field, as supported by various UCoAl experiments~\cite{Brando2016,Karube2012,Shimizu2015}. 
In this context, the critical exponent of the relaxation time is denoted as $z\nu$, where $z$ is the dynamic exponent and $\nu$ is the critical exponent of the correlation length. 
For the 3D Ising model, the values of $\nu$ = 0.630 and $z$ = 2.0235 yield $z\nu$ =1.27~\cite{Guillou1980,Adzhemyan2022}.
However, the observed value of 1in UCoAl differs from the prediction of the 3D Ising model and is closer to the classical value deduced from $\nu = 1/2$ and $z = 2$. 
This experimental finding strongly suggests that the elastic anomaly near the CEP is not a critical relaxation phenomenon associated with spins, but rather should be attributed to other degrees of freedom.

Next, let us discuss the ultrasonic dispersion phenomena in UCoAl. 
The observed relaxation time is unusually long for solids and is comparable to values reported for liquid systems such as nematic liquid crystals~\cite{Eden1973}. 
A critical slowing-down phenomenon is also observed near the magnetic phase transition, where the elastic constants exhibit a minimum and the ultrasonic attenuation increases.
Typically, relaxation times in solids are on the order of 10$^{-11}$ s. 
A similar case was reported in CeRu$_{2}$Si$_{2}$, which shows a metamagnetism at $H_{\text m}$ = 7.7 T~\cite{Yanagisawa2002}. 
Significant elastic softening and ultrasonic attenuation were observed at 550 mK with measuring frequencies of 75 and 45 MHz, and the relaxation time was estimated to be on the order of 10$^{-9}$ s. 
To the best of our knowledge, however, the long relaxation times reported here have never been observed in solids.
The relaxation time represents the timescale of the energy dissipation process. 
According to mode coupling theory, it characterizes the energy dissipation process from the strain-induced state to various modes of the surrounding environment. 
In the case of nematic liquids, the relaxation time is characterized by the rearrangement or reorientation time of the strained molecules. 
It is worth considering whether the observed long relaxation times in UCoAl can be explained by such a simple mechanism. 
However, it is likely that more exotic mechanisms are involved. 
If the paramagnetic ground state is highly anomalous and competes with the induced ferromagnetic state above the metamagnetic transition, the timescale of excitation and relaxation processes between the two states driven by elastic strain may be long. 
The observed ultrasonic dispersion phenomenon implies the existence of a peculiar and intriguing state in this system.

\subsection{Origin of Elastic Anomalies}
Indeed, based on the power-law analysis of $C_{33}$ and the estimated giant elastic anomaly in UCoAl, it is suggested that an unknown order parameter, possibly a quadrupole, is responsible for the exotic elastic properties. 
However, the information obtained solely from $C_{33}$ does not exclude the possibility of charge degrees of freedom, considering the hexagonal crystal symmetry of UCoAl.
It is worth noting that a large anisotropic volume change has been observed at the metamagnetic transition in UCoAl, which suggests the involvement of charge fluctuations. In the case of UTe$_2$, for example, charge fluctuations are known to play a role in the metamagnetic transition~\cite{Miyake2022}. 
However, in UTe$_2$, the valence fluctuation state exists in the paramagnetic region, and charge fluctuations are inherent to the system.
In contrast, there is currently no evidence of valence fluctuations in UCoAl. 
Therefore, it is speculated that a quadrupole is a plausible candidate for the unknown order parameter responsible for the observed elastic anomalies in UCoAl. 
Further investigations are required to confirm the nature of this order parameter and understand its role in the unique properties of UCoAl.

The observations from XMCD experiments, indicating the presence of both spin and orbital degrees of freedom in UCoAl, provide valuable insights into the nature of this system~\cite{Kucera2002,Combier2017,Takeda2018}. 
The magnetic moment carried by the orbitals is found to be larger than that of the spins, and the orbital magnetic moments also exhibit metamagnetic behavior. 
This suggests that orbital degrees of freedom play a significant role in the physics of UCoAl and behave differently from spins.
The interplay between spin and orbital degrees of freedom in UCoAl can be attributed to the dual character of itinerancy and localization of the $5f$ electrons. 
The non-centrosymmetric crystal structure of UCoAl further enhances the mixing of magnetic and electrical degrees of freedom, making the two degrees of freedom appear to cooperate.
The presence of orbital degrees of freedom, which have the same symmetry as the strain, can give rise to elastic anomalies. In the case of the Spin-1 Ising model with both short-range magnetic and quadrupolar interactions, quadrupolar phases can emerge in addition to TCP and magnetically ordered phases~\cite{Levitskii2006}. 
This suggests that the quadrupolar nature of the orbital degrees of freedom in UCoAl could be responsible for the observed elastic anomalies and the exotic properties near the metamagnetic transition.
Further investigations and theoretical modeling are needed to fully understand the role of orbital degrees of freedom and their interplay with spins in UCoAl and to elucidate the origin of the quadrupolar order parameter.

The observations of anomalous properties near the QCP in UCoAl and similar systems suggest the presence of unconventional ground states. 
The temperature dependence of the electrical resistance in UCoAl, exhibiting a critical exponent different from that of the 3D Ising model, indicates an anomalous ground state near the QCP~\cite{Maeda2018}. 
Similar behavior has been observed in UGe$_{2}$~\cite{Tateiwa2014}.
The large maximum in the magnetic susceptibility for $H \parallel c$-axis at $T_{\chi_{\rm Max}}\sim 20\,{\rm K}$ cannot be solely explained by ferromagnetic fluctuations. 
In UCoAl, the susceptibility and $1/(T_{1}T)$ (where $T_1$ is the spin-lattice relaxation time) exhibit a maximum at 20 K~\cite{Karube2012}. 
NMR experiments under uniaxial pressure reveal that ferromagnetic and TCP states emerge with increasing pressure along the $c$-axis, and the 20 K anomaly arises from TCP~\cite{Karube2014}. 
The anomaly observed in $C_{33}$ around 20 K and the anomaly released in the $C_{44}$ elastic constant~\cite{Kumano2020} may have the same origin as this magnetic susceptibility maximum.

The spin-spin relaxation rate $1/T_2$, which reflects longitudinal spin fluctuations, increases toward the CEP, indicating strong longitudinal fluctuations in UCoAl~\cite{Nohara2011,Karube2015}. 
The peak position of $1/T_{2}$ remains constant at around 10 K over a wide range of magnetic fields. 
A similar anomaly to $1/T_{2}$ is observed in $C_{33}$, as shown in Fig.~\ref{fig8}(a). 
From the CEP to zero magnetic field, the $C_{33}$ minimum is located parallel to $T_{\chi_{\rm Max} }$. 
The 12 MHz data in Fig.~\ref{fig8}(a) exhibit this anomaly above 0.55 T. 
These two anomalies are attributed to the same underlying mechanism.

Theoretical approaches based on band pictures have been employed to understand TCP in itinerant ferromagnetic materials~\cite{Yamada1993,Belitz2005}. 
The magnetic susceptibility maximum arising from TCP can be explained by both band theory and the spin-1 Ising model~\cite{Yamada1993,Levitskii2006}. 
Given that the characteristic energy scale of UCoAl is approximately one order of magnitude smaller than that of typical itinerant metamagnetic materials such as Co(S$_{1-x}$Se$_{x}$)$_{2}$~\cite{Goto1997}, 
it is reasonable to consider CEPs and TCP from a localized electron perspective based on the spin-1 Ising models, such as the Blume-Capel model and the Blume-Emery-Griffiths model~\cite{Blume1971,Jain1980,Levitskii2006,Yang2008,Ejima2020,Zivieri2022}.

The spin-1 Ising model can be mapped onto the transverse field spin-1/2 Ising model by replacing the role of spin-0 with a hole~\cite{Oitmaa2003,Yang2008}. I
n this model, a hole-condensed state appears as the ground state, and the ``large spin state'' tends to disappear at low temperatures under the influence of a uniaxial anisotropy.
The control of the dipole moment state by uniaxial stress suggests that strains and quadrupoles play a crucial role in the emergence of first-order phase transitions, 
TCP, and related phenomena. 
Considering the theoretical prediction of the 3D Ising model, we can infer that the spin magnitude decreases with temperature, and the anomalies observed in $1/(T_{1}T)$ and $1/T_{2}$ are crossovers related to the spin magnitude. 
These mysterious paramagnetic states reveal themselves through the peculiar elastic properties.

It is worth noting that UCoAl has a ZrNiAl-type hexagonal structure, where U atoms form a quasi-kagom\'{e} structure, implying the presence of possible magnetic frustrations and spin nematic instabilities. 
The spin-1 bilinear and bi-quadratic model, which extends the spin-1 Ising model, exhibits more pronounced features on triangular lattices~\cite{Lauchli2006,Remund2022}. 
In this model, the spin-0 states can be interpreted as quadrupole moments. 
The model predicts the emergence of ferro-quadrupole (FQ) and antiferro-quadrupole orders in addition to antiferromagnetic (AFM) and ferromagnetic (FM) orders, depending on the competing interactions. Spin nematic order is also expected to arise~\cite{Tsunetsugu2006}. 
On a triangular lattice, the FQ phase is located near the FM phase, and dipole and quadrupole phases can coexist~\cite{Remund2022}. 
By appropriately selecting the model parameters, a quadrupole phase can be hidden by the dipole phase~\cite{Remund2022}. 
In this scenario, a significant elastic anomaly is anticipated since the corresponding elastic strain possesses the same symmetry as the quadratic order parameter. 
The observed elastic softening can be attributed to the apparent or hidden quadrupolar order.
Here, we can draw a parallel with the case of the critical endpoint (CEP) between $\gamma$-Ce and $\alpha$-Ce, where an abrupt volume change occurs along with a softening of the bulk modulus. Although both phases retain a cubic system, the space group changes from $Fm\bar{3}m$ in $\gamma$-Ce to $Pn\bar{3}m$ or $Pa\bar{3}$ in $\alpha$-Ce. 
This phenomenon is known as a ``hidden'' structural phase transition, where the discussion revolves around quadrupole ordering without dipole moments~\cite{Nikolaev2012}.

The two scenarios we discussed, the nematic scenario based on the spin-1 bilinear and bi-quadratic model and the entanglement of dipole and quadrupole degrees of freedom, provide potential explanations for the cooperative phenomena observed in UCoAl. 
It is interesting to note that similar behaviors, where magnetic and strain susceptibilities diverge toward the same critical temperature, have also been observed in iron-based superconductors~\cite{ning2010,yoshizawa2012-1}.
This similarity suggests a common underlying physics in these systems~\cite{fernandes2010}.
The measure of the coupling between strains and quadrupole degrees of freedom, $E_{\text{JT}}$, obtained in the previous analysis in Sec.~\ref{Powerlaw}, provides valuable information for identifying the nature of the quadrupole. 
The small values of $E_{\text{JT}}$ compared with those observed in iron-based superconductors~\cite{yoshizawa2012-1} could serve as a clue in distinguishing the quadrupole in UCoAl. 
These phenomena remain highly puzzling and intriguing topics in the field of solid-state physics~\cite{Fernandes2014}.
In the context of related compounds, a recent study reported a nontrivial phase transition at 54 K in URhSn, a material isomorphic to UCoAl~\cite{Shimizu2020}. 
The authors of that study discussed the possibility of quadrupole ordering in URhSn, suggesting that the quadrupole degree of freedom plays an important role in this family of compounds. 
Our finding further highlights the significance of quadrupole effects near ferromagnetic quantum phase transitions in these uranium compounds.  

\section{Summary}
The discovery of elastic softening near the magnetic CEP in UCoAl is a striking feature of this study. 
It was unexpected and represents a significant finding. The elastic constant $C_{33}$ exhibits a pronounced anomaly in the vicinity of the CEP, indicating the presence of exotic elastic behavior. Importantly, the observed elastic anomaly is characterized by an extremely long relaxation time, the longest ever observed in solids. 
This suggests the presence of a huge static elastic anomaly associated with the quadrupole order parameter, which has the same symmetry as $\epsilon_{zz}$.
A model has been constructed to describe the temperature and magnetic field dependence of the relaxation time ($\tau$), which successfully reproduces the experimental data. 
The $\tau$ exhibits three distinct regions, with Region III showing a strong suppression of $\tau$ for some reason. 
One possible scenario is that $\tau$ is composed of two components: a magnetic field-dependent component and a less sensitive component. Investigating the properties of $\tau$ could provide further insights into the nature of the quadrupole.
The anomalous acoustic phonon dispersion near the $\Gamma$ point in the Brillouin zone, as revealed by the remarkable static softening of $C_{33}$ and ultrasonic dispersion, is another interesting feature. 
The acoustic phonon frequency $\omega$ is typically proportional to the wave number $k$, with the slope representing the sound velocity. However, in UCoAl, the results suggest the presence of higher-order terms, such as a $k^2$ term, in the longitudinal phonon frequency, which become significant near the CEP. 
The decrease in sound velocity and the unusually long relaxation time near the CEP point to an exotic ground state.
Two scenarios have been discussed to explain the origin of these anomalous elastic properties; nonetheless, further investigations are needed to fully understand these intriguing behaviors in UCoAl.

\begin{figure}
\centering
\includegraphics[width=15cm]{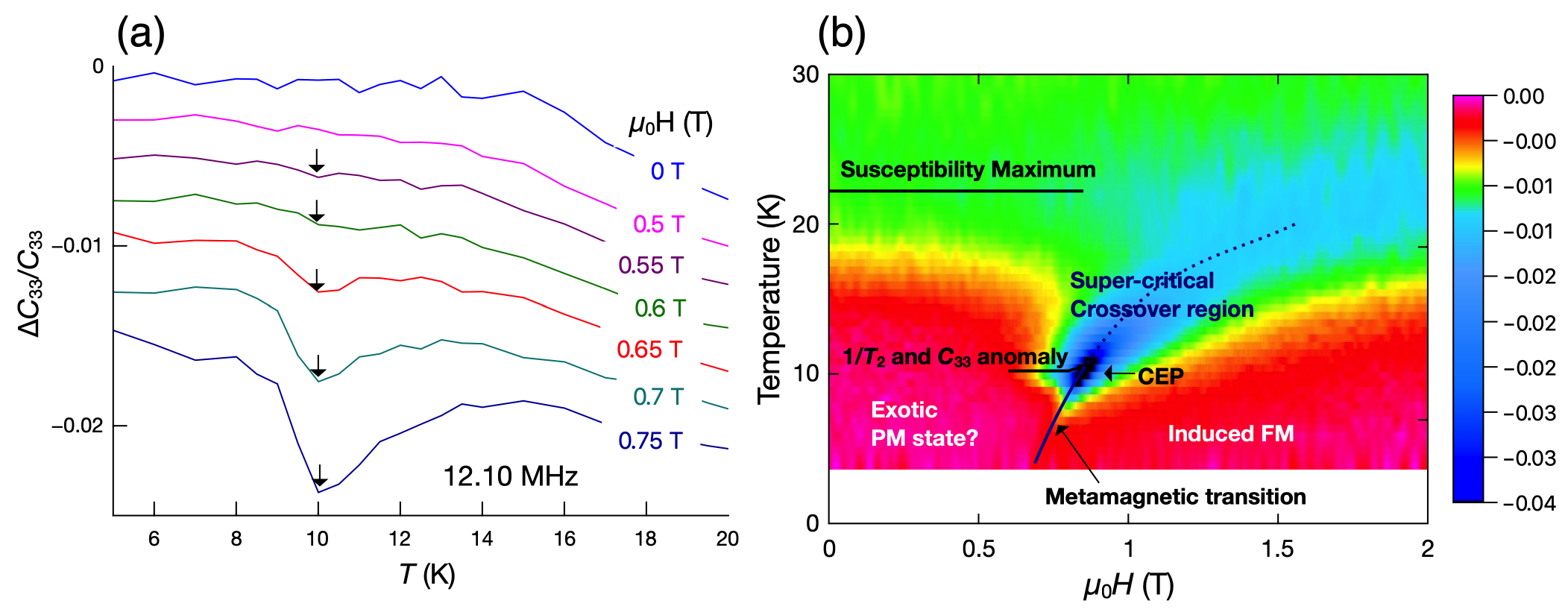}
\caption{
(Color online) (a) Temperature dependence of $C_{33}$ in selected magnetic fields, with minima around 10 K arising from the CEP field.
(b) Phase diagram of UCoAl color-coded by $C_{33}$ values. 
The metamagnetic transition (solid line) following the magnetic susceptibility maxima and supercritical lines (dotted line) found by magnetic measurements is well reflected in the $C_{33}$ profile.
The location of the CEP is also appropriately determined as a sharp $C_{33}$ anomaly.
}
\label{fig8}
\end{figure}

Figure~\ref{fig8}(b) presents a color-coded diagram representing the magnitude of the elastic constant anomaly, with lines indicating the positions of elastic constant minima. 
In addition to the previously known metamagnetic lines and the susceptibility maximum at 20 K, a new line with an elastic constant minimum extending from the CEP to zero field has been discovered. 
This new finding suggests the presence of an additional phase transition or critical phenomenon associated with the elastic properties of UCoAl.

In general, CEPs provide a unique and extreme environment for studying various physics, and their nature is still largely mysterious. 
They serve as a promising platform for investigating hidden aspects and uncovering novel and exotic physics. 
This study serves as a starting point for future research, and we anticipate that it will contribute to the exploration and understanding of the intriguing physics of CEPs.

\begin{acknowledgments}
We thank S. Goto (Lightom Corporation) for fabricating the piezoelectric transducers specifically for this study and M. Nakamura (Iwate University) for his assistance in setting up the experiments. 
We also thank Y. Tokunaga (Japan Atomic Energy Agency) and T. Yanagisawa (Hokkaido University) for providing useful information and valuable discussions.
Y. S. was supported by JSPS KAKENHI Grant Numbers JP20K03851, JP23K03314 and JP23H04870
Y. N. was supported by JSPS KAKENHI Grant Numbers JP18K03530 and JP21K04622.
\end{acknowledgments}

\bibliographystyle{jpsj}
\bibliography{Yoshizawa,Ultrasonics,UCoAls,FeAs}

\end{document}